# Adaptive Sparse Sampling for Quasiparticle Interference Imaging

*Jens Oppliger[1], Berk Zengin[1], Danyang Liu[1], Kevin Hauser[1,2], Catherine Witteveen[1,3], Fabian von Rohr[3] and Fabian Donat Natterer[1]\**

1. Department of Physics, University of Zurich, Winterthurerstrasse 190, CH-8057 Zurich, Switzerland
2. Department of Physics, Harvard University, 17 Oxford Street Cambridge, MA 02138, USA
3. Department of Quantum Matter Physics, University of Geneva, 24 Quai Ernest-Ansermet, CH-1211 Geneva, Switzerland

Quasiparticle interference imaging (QPI) offers insight into the band structure of quantum materials from the Fourier transform of local density of states (LDOS) maps. Their acquisition with a scanning tunneling microscope is traditionally tedious due to the large number of required measurements that may take several days to complete. The recent demonstration of sparse sampling for QPI imaging showed how the effective measurement time could be fundamentally reduced by only sampling a small and random subset of the total LDOS. However, the amount of required sub-sampling to faithfully recover the QPI image remained a recurring question. Here we introduce an adaptive sparse sampling (ASS) approach in which we gradually accumulate sparsely sampled LDOS measurements until a desired quality level is achieved via compressive sensing recovery. The iteratively measured random subset of the LDOS can be interleaved with regular topographic images that are used for image registry and drift correction. These reference topographies also allow to resume interrupted measurements to further improve the QPI quality. Our ASS approach is a convenient extension to quasiparticle interference imaging that should remove further hesitation in the implementation of sparse sampling mapping schemes.

**Method details**

Broader context

Sparse sampling for quasiparticle interference (QPI) imaging is a novel scanning tunneling microscopy mapping scheme that promises fundamentally shorter measurement times than conventional grid spectroscopy(*1*, *2*). The speed-up from sparse sampling comes from measuring far fewer samples than suggested by the Nyquist-Shannon sampling theorem. The condition for a successful compressed sensing[1] recovery(*3*, *4*) is a high degree of signal sparsity in some representation space. For QPI imaging, the Fourier transform of the local density of states (LDOS) can be highly sparse(*1*, *5*), which incidentally is also the representation space. When sparse sampling is paired with faster point-spectroscopy(*6*), QPI mapping becomes orders of magnitude faster than conventional grid methods(*2*). This leaves the question as to the required sub-sampling as an unresolved practical issue because the mathematical prescriptions that determine successful recovery(*7*) depend on *a-priori* insight about the signal-to-noise level and the initially unknown degree of signal sparsity.

Here we introduce an adaptive sparse sampling (ASS) approach through which we iteratively increase the amount of sub-sampling during runtime to accumulate LDOS measurements until a satisfactory signal quality becomes visible in the preview snapshots of the recovered QPI images. Our ASS procedure further enables an interruption of QPI imaging at recurring exit points from which the mapping can be resumed to increase the cumulative amount of LDOS measurements. The interleaving of regular topographic scans with the sparsely sampled LDOS measurements enables image registry and drift correction that may be required due to thermal drift or after repositioning of the tip in the aftermath of longer interruptions, such as the refilling of one's cryostat. Through ASS, an experimenter can design open-ended mapping tasks and tackle the QPI inspection of systems for which there is only incomplete information available about the sparsity level and noise. Our ASS method allows to monitor the signal quality in-operando for feedback of ongoing measurements and to stop/interrupt measurements at opportune moments.

---

[1] Sparse sampling, compressive sensing, and compressed sensing are used synonymously.

Method detail

The working principle of our ASS approach is summarized in Figure 1, where we first exemplify the concept using a simulated Shockley surface state consisting of a standing wave modulating the LDOS, atomic corrugation, and added Gaussian noise. The full LDOS and QPI, shown here to the left for one energy, is the desired information (ground truth) we seek to recover. The scattered surface state is modelled using a modified Bessel function of the first kind(8). The random motion path overlay indicates the traveling salesperson route for one sparse sampling measurement and the dashed box marks the reference topography (or LDOS) that is used for image registry, as described below. To approach the QPI ground-truth of the surface state ring and the Bragg peaks in an adaptive measurement scheme by gradually accumulating LDOS measurements, we proceed according to the following four steps:

1. Path generation
2. Path combination
3. Mapping and preview
4. Postprocessing corrections.

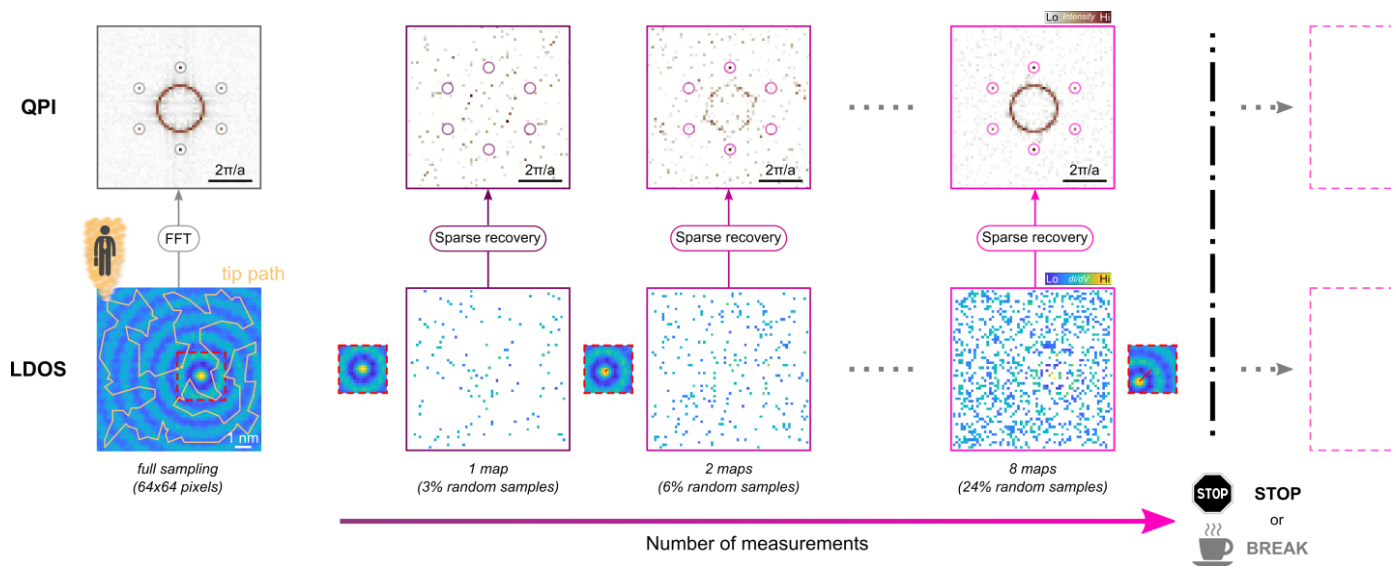

*Figure 1: **Adaptive sparse sampling concept.** The top panels show the quasiparticle interference (QPI) patterns of a simulated surface state. The leftmost QPI pattern is the ground truth that we want to adaptively reconstruct from an accumulation of sparsely sampled local density of states (LDOS) measurements. The reconstruction gets gradually better with increasing number of cumulative measurements. The bottom panels show the LDOS with the ground truth in the leftmost panel. The orange line indicates the traveling salesperson path that is used to obtain the sparsely sampled LDOS measurements in the adaptive sampling scheme. The small insets between the maps illustrate the usage of smaller LDOS or topographic maps that can be intersected between adaptive sampling measurements to align the individual measurements for the combined sparse recovery and the red line indicates the drift vector. In an actual measurement one should use regular topography scans due to the much shorter acquisition time. The ASS concept allows to pause, resume, or interrupt the measurements, for instance when the QPI pattern has a sufficient quality.*

1. **Path generation**

Prior to the actual measurement, we prepare the tip travel paths. From all possible locations on a grid of size $N = n \times n$, we select $p$-times $qN$ random locations with $q \in (0,1]$ being a small fraction. For every subset $p$, we calculate a near optimal solution to the traveling salesperson path (TSP) using a genetic(9) or cheapest-insertion(10) algorithm with fixed start and end coordinates. These two fixed points are shared among all paths to simplify their connection between the individual iterations, and they help build a catalogue of precalculated paths that can be quickly combined in the preparation of an actual measurement. Here we limit ourselves to a maximal number of $qN = 3000$ locations per path because of the poor scaling of TSP calculations (computational NP-hardness). Note that the choice of these locations is compatible with informed sampling(1) in which we modify the probability of selecting a location based on various criteria, such as: proximity to the border, no-go areas, or defects of high LDOS scattering intensity. The average time for calculating one TSP path segment containing 3000 locations is about 2.5 hours per CPU core (AMD EPYC 7302P) using a genetic algorithm. The path generation using a cheapest insertion algorithm is faster but results in a slightly longer overall path length. We precalculated $\sum p = 100$ TSP paths of which we typically use 20 to 30 in the present demonstration, equivalent to 60'000 to 90'000 locations at which the full LDOS is measured.

**2. Path combination**

The experiment starts by measuring a regular topography, preferably centered around topographic features that may serve as fiducial markers to aid image registry. We then proceed to set up the path for an ASS experiment where we define the number of TSP path subsets that we measure consecutively before we scan a regular topography. The reason for enchaining several TSP segments is the limited number of locations that we can afford to compute for one path and the relatively long duration of the topographic scans in our early ASS implementation, which consisted of measuring a full spectrum also for every location in the reference topography. Since full spectroscopic information as reference for the topography is not necessary, a regular topography scan would be sufficient and much faster. The end of the topography scan concludes the first ASS iteration. Since salient topographic features for image registry may be off-center, we generate a jittery random-walk path from the last location of the TSP segment to the start location of the topography scan and another jittery path from the last location of the topography scan to the start of the next TSP subset. This allows the system to settle creep that may arise from larger changes in the scan piezo voltage from repositioning the tip. By default, the ASS routine then proceeds to the next path subsets and records another topography scan at its end and so forth. The correlation between the reference-topographic scans yields the effective displacement that occurred from one iteration to the next due to thermal drift or residual creep. This displacement information is used at the end of the QPI mapping to accurately assign the correct coordinates to each location, which improves the overall reconstruction quality, as discussed further below.

**3. Mapping and preview**

We obtain the QPI information indirectly by first measuring the LDOS at every of the chosen locations, as described by the previously generated path combination, and then perform a sparse recovery of the QPI image by solving a basis pursuit denoising problem provided by the sparse least-squares solver SPGL1(*11*, *12*) written in MATLAB. The procedure to recover the QPI pattern from LDOS measurements is equivalent to our previous sparse sampling implementations(*1*, *2*) and does not depend on the ASS method per se. The only difference is that our measurement matrix that feeds the sparse sampling recovery solver is growing with every iteration; see supplementary materials (*SM*) for a brief mathematical description. We measure the LDOS either via conventional bias spectroscopy or via parallel spectroscopy(*2*) for several energies. All LDOS measurements of one energy are treated independently to the LDOS at a different energy. In addition to the spectral information, we also record the time at which each spectrum is saved. This timestamp could later serve for displacement correction(*1*). The series of panels in Figure 1 shows how more and more LDOS measurements (bottom row) are accumulated after every TSP path iteration, leading to a growing measurement matrix. The top-row shows the QPI image that is obtained from the sparse recovery of the cumulative LDOS. We can appreciate how the quality of the recovered QPI pattern improves with more sampling. To use this mapping scheme in an actual measurement, we limit ourselves to the reconstruction of a few selected energies during the regular topography scans. The sparse recovery computation of the QPI pattern from a $1024 \times 1024$ grid for a single energy slice takes only a few minutes, leaving sufficient time to decide whether to proceed with the mapping or to stop. This preview option is useful to gauge the quality of the QPI mapping and to get a sense for the level of detail already present in the QPI maps.

**4. Postprocessing corrections**

**(a) Accounting for tip instabilities:** A frequent disturbance in STM investigations are slight and spontaneous tip-changes that manifest as an abrupt change in the conductance. To salvage such measurements, we perform a global background correction that conserves both the long-range LDOS modulations and the LDOS relationship between the measurement locations. Consequently, these background corrections preserve the spatial frequency content in the LDOS from which the QPI signature is obtained. We assume that the measured conductance spectra are a convolution of tip ($\varrho_T$) and sample ($\varrho_S$) density of states ($dI/dV(E) \propto \varrho_S \varrho_T$)(*13*). If $\varrho_T$ changes, this will merely have an energy dependent multiplicative effect on the conductance. We also assume that we start our measurements or are able to identify segments with a well-behaved tip that has finite conductance in the relevant energy interval. The conductances measured for different energies/bias voltages are treated independently. In order to combine the LDOS data, measured for different TSP segments and with different tip density of states $\varrho_T$, we proceed as follows: We first calculate for every TSP segment $s$ the respective mean $m_s = \sum_k \varrho_S^k \varrho_T^k / qN$ and standard deviation $\sigma_s = \sqrt{\sum_k (\varrho_S^k \varrho_T^k - m_s)^2 / qN}$ independently for all bias voltages and locations $k$ within that TSP segment. We then select one TSP segment as reference (preferably one without tip-changes, which typically is the first one) and subtract the individual mean from all other TSP segments, divide by their respective standard deviation, multiply by the standard deviation of the reference, and add the mean of the reference, $\varrho_S\varrho_T' = (\varrho_S\varrho_T - m_s)\sigma_{ref}/\sigma_s + m_{ref}$. We apply this correction to TSP segments that show no tip-changes during their measurement, although such data could be handled in a similar way. One could identify the measurement index at which the tip-change occurred using an edge-detector and then treat one sub-segment as reference for the remaining sub-segments with the correction just described. Note that a different tip configuration might have a detrimental impact on the spatial frequency content due to its spatial convolution that inhibits the resolution of high-momenta states. Such states are however not relevant to our present demonstration.

**(b) Linear drift correction:** As mentioned, the interleaved topographic images serve to create spatial references for the TSP path segments such that the LDOS measurements of distinct ASS iterations can be combined into one measurement matrix for QPI reconstruction. Depending on the stability of the system or whether the measurement has been interrupted for some time, the image registry can become essential to ensure the proper spatial relationship between LDOS measurements, notably between those of distinct TSP subsets. The knowledge of the exact coordinates is what enables the sparse recovery for QPI imaging. To account for the time-dependence between the topography images and the different distances between consecutive locations, we can use the timestamp information $t_k$ of every location $k$ that we have recorded alongside the LDOS (*SM*). Following our previous work(*1*), we first determine the global displacement vector $\vec{v} = (v_x, v_y)$ from the image registry between reference topographies and the drift-speed $d\vec{v}/dt$, calculated from the total time that has passed in-between the reference images. We assume linear displacement and accordingly attribute to every location a proportional correction $\vec{r}_i' = \vec{r}_i + \frac{d\vec{v}}{dt}(t_k - t_{k-1})$. This adjusts the effective coordinates of the LDOS measurements, which helps in the QPI reconstruction because it improves the reliability of the spatial relationship in the LDOS modulations. In the (*SM*), we discuss nonlinear drift and scenarios where these assumptions cease to apply.

Method validation

In order to validate the functioning for an actual measurement, we choose the well-known model system Au(111), which is characterized by a Shockley surface state and a $(22 \times \sqrt{3})$ herringbone reconstruction(*14*). In our previous work(*2*), we have measured the dispersion relation of the nearly free electron gas using all combinations of conventional spectroscopy, parallel spectroscopy and sparse sampling for QPI imaging to ensure that our spectroscopies and mapping modes reproduce the Au(111) reference signatures. However, in those measurements we had measured all LDOS values in a single sweep, that is, not adaptively. We proceed here to reproduce these characteristic Au(111) signatures using the adaptive sparse sampling implementation. For the present ASS demonstration, one iteration consists of 5 random walk segments with 3'000 measurement locations each followed by one reference topography. The first iteration is started with an additional topography. Since a reference topography requires about 20 minutes, it can also be interpreted as an interruption/break of the sparse sampling mapping scheme. The panels (a)-(c) in Figure 2 show the evolution of the QPI reconstruction with an ever-growing number of adaptively sampled LDOS values that are added to the reconstruction with the applied background correction mentioned previously in the post-processing section. From the dispersion plot in panel (a), one clearly notices how more band-structure details emerge with increasing number of measurements. We also see that the states closer to the reciprocal zone center appear at lower sampling than high momentum-transfer states, which is related to the effective level of sparsity at the respective energies in *q*-space. When the surface state has a higher *q*-value, it is represented by more wavevector values simply due to the larger circumference(*1*). Similarly, the herringbone reconstruction is visible from a rather low sampling already, showing the dependence of the reconstruction efficiency with sparsity in *q*-space. We further elaborate on this in the discussion below. The inverse Fourier transforms of the QPI images are shown in panels (c) and reveal the spatially resolved LDOS with standing wave patterns and the surface reconstruction. For comparison, we also show the same data without background correction in panel (d). The dispersion plots are perturbed by sudden jumps that are caused by tip-changes, occurring during our measurements. This can also be seen in Figure 2(e), where we show how the background correction accounts for and removes sudden changes in the conductance. As mentioned above, this correction is benign with regards to the spatial wavevector information. Although of lesser quality, the uncorrected QPI reconstruction does still show an improvement with increasing ASS iterations, which is crucial for quality assessment during runtime.

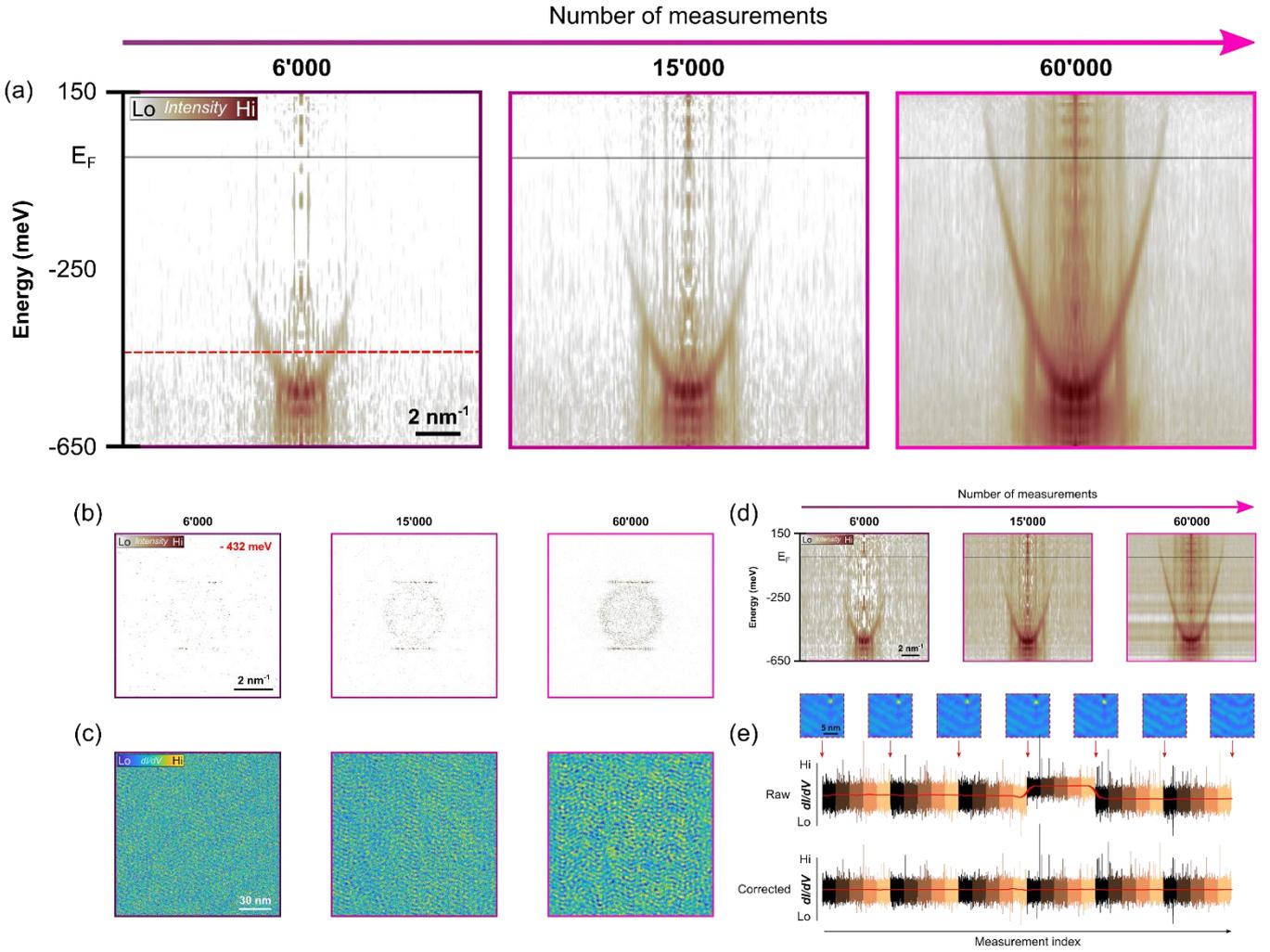

*Figure 2: **Validation of adaptive sparse sampling using Au(111).** (a) Dispersion plots of Au(111), showing the parabolic dispersion of the nearly-free electron like Shockley surface state. A background correction, as described in the text, has been applied and the dispersion plots are created from azimuthal averages of QPI patterns, such as the ones shown in (b). The quality improves with increasing number of adaptively sampled local density of states (LDOS) measurements. **(b)** QPI patterns obtained after sparse recovery of adaptively sampled LDOS, showing the gradual improving quality with increasing number of measurements. **(c)** LDOS obtained from an inverse Fourier transform of the QPI patterns in (b). Every ASS increment consisted of 3'000 locations and a reference topography/LDOS was recorded after 5 such ASS increments. **(d)** Same dispersion plots as in (a) but without background correction applied, showing the detrimental effect of tip-changes. **(e)** Reference topographies/LDOS maps that are intersected between the adaptive sparse sampling loop (interleaving shown by vertical red arrows) to align the LDOS measurements for the cumulative sparse reconstruction. The two bottom rows show an example of the LDOS trace before and after the application of our background correction. (setpoint: $V_b$ = -250 mV, $V_{drv}$ = 400 mV, $f_{drv}$ = 1600 Hz, $I_t$ = 1.5 nA, $t_{spc}$ = 20 ms, T = 4.3 K, grid size 1024x1024).*

Discussion und tweaks

We have demonstrated the mechanism through which QPI measurements can be adaptively and sparsely sampled. We now turn to discussing how to generate accompanying quantitative feedback that could serve as a quality assessment for the QPI data. To that end, we calculate for every ASS step the relative mean absolute error (MAE) *(SM)* between the cumulative measurements with and without the latest iteration, which can be done at runtime. As shown Figure 3(a) we notice a rapidly decreasing MAE with progressing ASS iterations, indicating the gradual approaching of the ground truth. The insets in (a) relate this quantitative measure to the more familiar QPI pattern for visual quality assessment. This indicates that the MAE could serve as a useful indicator for the QPI quality, enabling the numerical tracking of the QPI progress in automated decision-making approaches such as reinforcement learning. In panel (b), we apply the MAE evaluation also to our actual QPI measurement of the Au(111) surface state. The MAE likewise nicely tracks the quality of the QPI reconstruction. We further notice a dependence on the LDOS energy with regards to MAE improvement, which can be traced back to the varying degree of sparsity in the QPI pattern for different energies. This can be seen when comparing the MAE for -551, -395, and 67 meV. The momentum of the energetically more positive surface state is higher, which translates into a lower sparsity because the surface state consists of more values due to its larger circumference.

Up to this point, we have mostly assumed well-behaved systems with a handful of smaller tip-changes for which we have introduced a mitigating background correction. We obtain less reliable results when nonlinearities start to become prominent, for instance in nonlinear thermal drift, or nonlinear responses of the piezo-actuators to fast tip-speeds or larger step sizes that produce signatures of creep and hysteresis. Procedures on how to account for those piezo related non-linearities are focus of a forthcoming work (Hauser et al.). In the (*SM*), we show the breakdown of the linear drift assumptions when the time between two consecutive reference topographies becomes too long.

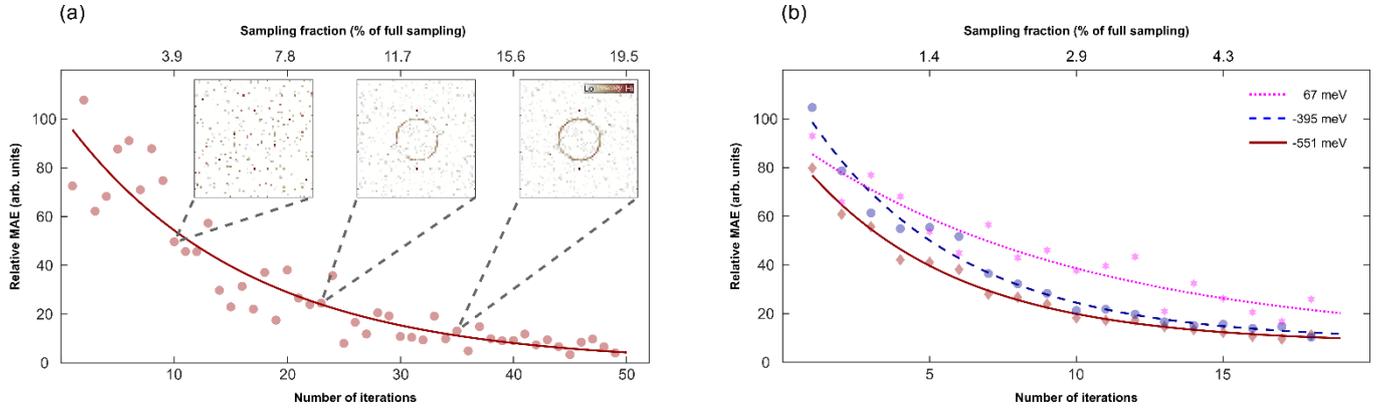

*Figure 3: **Quality of quasiparticle interference reconstruction with increasing sampling fraction.** **(a)** The decreasing relative mean absolute error (MAE) (SM) between two consecutive ASS iterations shows how the quality of the QPI reconstruction rapidly increases with growing number of iterations, here demonstrated for the simulated surface state of Figure 1 with known ground-truth. **(b)** The relative MAE for the actual QPI measurement on Au(111) shown in Figure 2 for three different energies, follows a similar trend. Since the relative MAE can be calculated after every ASS iteration at runtime, it provides feedback on the status of the QPI data and enables an informed decision about how much more sampling would be required. The different slopes in the three curves reflect the reduced sparsity at higher energies that requires appropriately more sampling. Note that the data in both (a) and (b) follow a power law behavior as indicated by the fitted lines.*

**Acknowledgements**

F.D.N. thanks the Swiss National Science Foundation (PP00P2_176866 and 200021_200639) and ONR (N00014-20-1-2352) for generous support. C.W. and F.O.vR acknowledge the support from the Swiss National Science Foundation (PCEFP2_194183) and by the Swedish Research Council (VR) through a neutron project grant Dnr. 2016-06955. We appreciate fruitful discussions with Jenny Hoffman and Jose Lado.

**Author contributions**

J.O. and F.D.N. conceived the project. F.D.N. and J.O. wrote the manuscript. J.O., and F.D.N. analyzed the data. B.Z., D.L., K.H. and F.D.N. measured the data. B.Z. D.L., and K.H. prepared the samples. C.W. and F.O.vR. synthesized the NbSe$_2$ samples. F.D.N. supervised the project.

**Supplementary materials and additional information**

Sample preparation and spectroscopy

We prepare clean surfaces of Au(111) by repeated cycles of ion bombardment (Ar$^+$ ions, 1.6×10$^{-5}$ mbar, 25', 4 µA/cm$^2$) and prolonged annealing at 620 °C. We perform our QPI measurements on terraces exceeding 200 nm. We use an electrochemically etched W-tip that we gently plunge into the Au sample until the apex is atomically sharp, which we verify by scanning across Au step edges. The individual spectra of the main text are recorded using the parallel spectroscopy technique, described earlier(*2, 6*). High quality single crystals of 2H-NbSe$_2$ were grown by the chemical vapour transport (CVT) method using I$_2$ as a transport agent. Stoichiometric amounts of Niobium (powder, Alfa Aesar, 325 mesh, 99.99%) and Selenium (shots, ground before use, Alfa Aesar, 99.999%) of a total mass of 0.5 g were sealed in a quartz ampoule (l = 15 cm, øl = 9 mm) under vacuum together with 45 mg of I$_2$ (Fluka, >99%). The ampoule was heated a rate of 180 °C/h in a two zone furnace in which the source zone was fixed at 850 °C and kept at this temperature for 6 days. Eventually, the tube was quenched in cold water. The product was confirmed to be phase pure by powder X-ray diffraction (PXRD). Patterns were collected on an STOE STADI P diffractometer in transmission mode equipped with a Ge-monochromator using Cu K$_{\alpha 1}$ radiation and on a Rigaku SmartLab in reflection mode using Cu K$_\alpha$ radiation. A single crystal of NbSe$_2$ is glued together with a cleaving post on top onto the sample holder, transferred without any further heat-treatment into our ultra-high vacuum system, and cleaved in-situ prior to transferring into the cooled scanning tunneling microscope. The spectra of NbSe$_2$ shown below in Figure 4(e) were measured with conventional lock-in spectroscopy.

Time stamp information

For every LDOS location, we measure the timestamp at which the data is recorded, which can be used for drift correction as described in the main text and in our previous work(*1*). Here we mention the two methods we have used to obtain timestamps. The more convenient is to use an event-logger (Swabian Instruments, TimeTagger Ultra), which directly provides a unique timestamp for every measurement location. A more accessible alternative and equally suitable consists in digitizing a periodic sawtooth waveform from an arbitrary waveform generator for time-information. The period should capture several measurement locations to ensure sufficient constraint as to the period number of the sawtooth waveform train. The time-information can then be calculated from the measured voltage level and the period number.

Sparse recovery and measurement matrix

The measured conductance values are coupled to the corresponding locations *k*. The general task of the sparse recovery in this case is solving the basis pursuit denoise problem given as $min\,||x||_1\,s.t.\,||Ax - y||_2 \leq \sigma$ where $A \in \mathbb{C}^{qN \times N}$ is the measurement matrix, $x \in \mathbb{C}^{N \times 1}$ is the solution to be reconstructed (in its sparse domain, i.e. (complex) Fourier space) and $y \in \mathbb{R}^{qN \times 1}$ includes the measured data in real space. $\sigma \in \mathbb{R}$ is a small error threshold that we set in terms of the standard deviation of the measured signal $y$. We construct the implicit measurement matrix $A = \Omega \Psi$ as a product of a restriction operator $\Omega \in \mathbb{R}^{qN \times N}$ and a 2D normalized fast Fourier transform operator $\Psi \in \mathbb{C}^{qN \times N}$. $\Omega$ is built such that the row *i* has a single "1" in *i*-th column *k(i)* and thus selects the entries listed in *k* from an input vector of length *N*. These operators are provided within the scope of the SPARCO toolbox(*15*). The measurement matrix $A$ and the measurement vector $y$ are then fed into the sparse recovery algorithm SPGL1(*11, 12*) along with the error threshold $\sigma$. By consecutively increasing the number of measurements, the matrix $A$ will have a higher number of rows, which means that more equations are used, leading to a more faithful reconstruction.

Relative mean absolute error (MAE) as feedback for reconstruction quality

We calculate the relative mean absolute error (MAE) of an image (in our case this is the reconstructed QPI image) with *n* rows and *m* columns as $\frac{1}{nm}\sum_{i=1}^{n}\sum_{j=1}^{m}\|\hat{X}_{ij} - X_{ij}\|_1/(X_{ij} + \varepsilon)$ where $\hat{X}_{ij}$ and $X_{ij}$ represent the pixel values in the *i*-th row and *j*-th column of the most recent and the previous cumulative ASS snapshots and $\varepsilon > 0$ is a small constant for numerical stability. A drop in the relative MAE over consecutive ASS iterations indicates a gradually approaching ground truth, with smaller gain in QPI quality per iteration.

Limitations of linear drift correction

In our previous work we introduced a linear drift correction procedure(*1*). The assumption of linear drift might not be adequate when there are larger thermal gradients in the system or when the duration between reference topographies is too long. In those situations, our linear drift correction would not improve the spatial relationship between individual measurements. An example in which our linear drift correction of nonlinear drift worsened the attribution of high-momentum states in NbSe$_2$ is shown in Figure 4. Panels (a) and (b) show the reconstructed quasiparticle interference patterns of NbSe$_2$ at -100 meV that we obtained on a $50 \times 50$ nm$^2$ region using 6'000 adaptively sampled points, corresponding to two individual TSP paths of 3'000 points each. We recorded the spectra using conventional d$I$/d$V$ spectroscopy taking 484 ms per spectrum. The grid resolution is 512×512 pixel. Before measuring the spectrum, we wait about 4.5 seconds to settle piezo-electric creep from the previous move to see the impact of thermal drift more clearly.

The main feature that is reconstructed in our ASS sparse recovery are the Bragg peaks forming a hexagonal pattern. Figure 4(b) shows the energy resolved momenta of NbSe$_2$ after applying linear drift correction. From the zoomed inset we see that the correction did not improve the sharpness of the Bragg peak and no dispersing electron states appear either. The absence of the dispersive electron states is attributed to the low-sampling rate. As mentioned in the main text, we use interleaved d$I$/d$V$ scans on a fully sampled sub-grid to extract the total drift distances as shown in Figure 4(c), which separates the ASS segments by 4.5 hours. We then utilize the simultaneously measured time stamps to account for an assumed linear drift to determine the actual locations of the measurements via post-processing. We observe that the applied linear drift correction leads to a worse reconstruction than without any correction as the region around the Bragg peak becomes more smeared. This indicates that our assumption of a purely linear drift does not apply, which we attribute to the hourlong delay between the two ASS segments, as seen by the provided time-stamps shown in Figure 4(c).

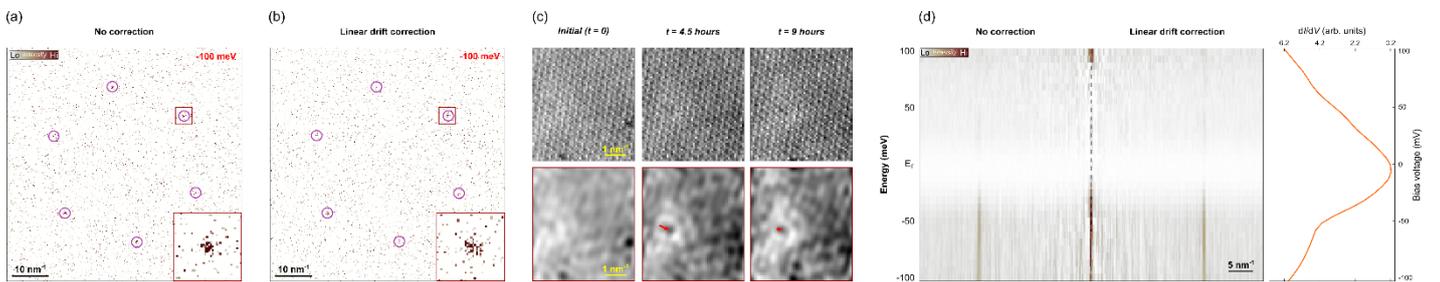

*Figure 4: **Adaptive sparse sampling applied to NbSe$_2$ illustrating the limit of linear drift correction.** (a) QPI pattern at -100 meV using two adaptively sampled paths consisting of 3'000 measurement points each. No correction has been applied to the locations of the measurement points. (b) Same as (a) but with a linear drift correction applied that we measured from topographic reference measurements in (c). The Bragg peaks appear more smeared, indicating the detrimental effect of nonlinear drift. (c) Topographic scans (top row) and low pass filtered topographies (bottom row) that we used to determine the drift vectors (in red) for the linear drift correction in (b). The corresponding time tag of the acquisition of each topographic scan is given on top. (d) Energy dependent momentum relation of NbSe$_2$ obtained using a 360-degree azimuthal average, showing the main Bragg peaks but no actual dispersive features, reminiscent of too low sampling rate to ensure their successful reconstruction. The average spectrum of NbSe$_2$ is shown for reference to the right.*

We expect a deviation from a truly linear thermal drift to have less influence on long-range real space features corresponding to low-momentum states in the scattering space as it is the case for Au(111). To investigate this statement, we show in Figure 5 a QPI simulation that we deliberately corrupt by drift. We study what impact our knowledge of the actual drift vectors or the presence of nonlinear drift have on the quality of the QPI pattern when we apply our linear drift correction. We simulate a Shockley surface state with underlying hexagonal atomic lattice as shown in Figure 5(a). In an adaptive sparse sampling fashion, we measure three times 1'000 unique points from a 256×256 pixel grid. After each iteration we shift the underlying ASS scan frame by a small random vector that is proportionally applied to the individual measurements, emulating the effect of a time-dependent thermal drift of the sample. The top panel in Figure 5(a) shows the reconstruction under the correct assumptions of purely linear drift and exact knowledge of the drift vectors. This corresponds to the best result that can be obtained. The lower two panels in Figure 5(a) illustrate what happens if we improperly measure the drift vector and are perturbed by linear drift or if we properly measure the drift vector but the measurements are perturbed by nonlinear drift. The QPI is most impacted in the lowest panel while the middle still shows a fair overall reconstruction. These singular examples are generalized in Figure 5(b)-(d), where we show simulations for different combinations of linear, nonlinear, and improperly measured drift vectors that we quantify using the multiscale structural similarity (MS-SSIM) (*16*) between the reconstructed output and the ground truth (Figure 5(a) top panel). A value of 1 for the MS-SSIM means identical images whereas lower values indicate poorer agreement between the recovered QPI and the ground truth. Our simulations show the phase space for good correction (dark blue regions) becomes small when nonlinear drift is coupled with improperly measured drift. These plots also emphasize the value of operating in the linear drift regime, which can be achieved by frequent interleaving of reference topographies or using fast single point spectroscopies.

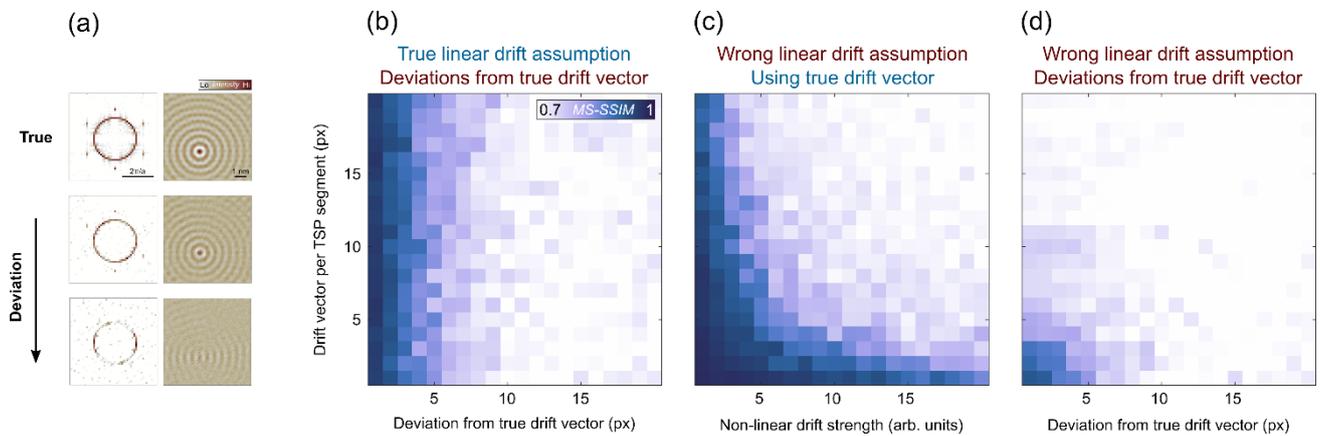

Figure 5: **Limit of linear drift correction due to wrong drift vector determination or nonlinear drift. (a)** Reconstruction of a simulated surface state with linear drift and exact drift-vector determination between each TSP segment consisting of 3x 1'000 points on a 256×256 pixel grid (top panel). The linear drift correction gets worse when the drift vector of a linear drift is improperly measured (middle panel) and encountering nonlinear drift while applying a linear drift correction leads to a loss of QPI features. While the Bragg peaks vanish very early, the surface state is more robust to drift due to its overall shorter scattering wavevectors (larger real space modulations). **(b)** Averaged simulation over five runs, assuming linear drift but improperly measuring drift vector between each TSP segment (both x and y direction), evaluated using the multiscale structural similarity (MS-SSIM) between the reconstructed surface state and the ground truth. We notice a continuously decreasing quality of the reconstruction with growing error in the drift-vector determination. **(c)** Correctly measuring the drift vector, correcting for linear drift, but encountering nonlinear drift between each TSP segment. The nonlinear drift is modeled using a quadratic Bézier curve. More nonlinear drift results in a more rapid degradation of the reconstructed QPI pattern. **(d)** Combination of (b) and (c), that is, erroneously measuring the drift vector and encountering nonlinear drift. The quality of the reconstruction worsens rapidly already for small errors in drift vector assignment and nonlinear drift.